\documentclass[conference]{IEEEtran}
\usepackage{graphicx}
\usepackage{amsmath}
\usepackage{amssymb}
\usepackage{cite}
\usepackage{hyperref}
\usepackage{amsmath, amsthm, amsfonts}
\usepackage{graphicx}
\usepackage{hyperref}
\usepackage{listings}
\usepackage{tikz}
\usepackage{color}
\usepackage{geometry}
\usepackage{array}
\usetikzlibrary{arrows.meta, positioning}
\usepackage{graphicx}
\usepackage{amssymb}
\usepackage{pifont}
\newcommand{\cmark}{\ding{51}} 
\newcommand{\xmark}{\ding{55}} 
\usepackage{pgfplots}
\usepackage{float}

\usepackage{pgfplotstable}
\usetikzlibrary{patterns, shapes, arrows.meta, positioning}

\usepackage{tikz}
\usepackage{pgfplots}
\usetikzlibrary{arrows.meta, shapes.geometric}
\pgfplotsset{compat=1.18}

\usepackage{pgfplots}
\usetikzlibrary{plotmarks}
\pgfplotsset{compat=1.18}

\usepackage{tikz}
\usetikzlibrary{calc}

\title{Rethinking Denial-of-Service: A Conditional Taxonomy Unifying Availability and Sustainability Threats}

\author{\IEEEauthorblockN{Mark Dorsett, Scott Mann, Tim Koussas}
\IEEEauthorblockA{\textit{La Trobe University} \\
\{M.Dorsett, S.Mann, T.Koussas\}@latrobe.edu.au}}

\begin{document}
\maketitle

\begin{abstract}
This paper proposes a unified, condition-based framework for classifying both legacy and cloud-era denial-of-service (DoS) attacks. The framework comprises three interrelated models: a formal conditional tree taxonomy, a hierarchical lattice structure based on order theory, and a conceptual Venn diagram. At its core, the taxonomy introduces six observable conditions (C0–C5) grounded in real-world attack behaviours, including source distribution, traffic volume, infrastructure targeting, and financial exploitation. These conditions enable consistent classification of known attacks—such as DoS, DDoS, LDoS, LDDoS, EDoS, DoW, and DDoW, while supporting identification of emerging or hybrid variants.

The lattice structure captures the cumulative satisfaction of conditions, allowing hierarchical reasoning across denial attack classes. The Venn diagram highlights conceptual overlaps between availability- and sustainability-focused attacks, improving comparative insight. Together, these models provide a robust analytical lens for threat modeling, mitigation strategy design, and attacker intent classification. The framework is particularly relevant in cloud-native and serverless environments, where sustainability-based attacks are increasingly impactful yet under-recognised. Its extensibility also permits future integration of socio-technical or behavioral dimensions.

By offering a structured taxonomy with theoretical grounding and real-world applicability, this work advances denial attack comprehension and equips defenders, researchers, and cloud architects with a shared vocabulary for interpreting and mitigating evolving threat vectors.\\
\end{abstract}


\section{Introduction}
The Denial Attack Conditional Taxonomy presents a novel and systematic framework for classifying a broad spectrum of popular denial attacks based on a set of six conditional characteristics, labelled C0 through C5. This taxonomy aims to unify legacy forms of denial attacks with more contemporary, economically-driven variants under a shared analytical structure. It departs from the conventional approach of categorising attacks merely by outcome, such as service unavailability or performance degradation, and instead emphasises the logical preconditions and behavioural markers that define the operational nature of each attack class. This enables a more granular understanding of attack mechanisms and offers a foundation for comparative analysis, threat modelling, and tailored mitigation design.

\section{Denial Attack History}

Denial attacks have evolved significantly over the past five decades, transitioning from rudimentary single sourced network flood tactics to sophisticated, economically motivated cloud-native vectors. Understanding the historical progression of these attacks is crucial to contextualising their classification under the proposed conditional taxonomy. This section provides a concise review of each major denial attack class, with emphasis on its defining mechanics and historical relevance.

\subsection{Denial of Service (DoS)}
The traditional DoS attack is characterised by the result of any action, event, or condition that limits legitimate access to, or the use of, a particular computing system resource \cite{menkus1997understanding}. This is performed by a single source overwhelming a victim service with excessive traffic or resource requests. These attacks often exploit protocol or application-layer weaknesses to crash systems or exhaust resources. Early examples include Ping of Death \cite{yihunie2018ping} and Teardrop attacks \cite{trabelsi2013hands} in the 1990s, which targeted OS-level vulnerabilities. While less common today due to improved system hardening, DoS attacks remain foundational to the denial threat spectrum and are primarily associated with availability disruption.

\subsection{Distributed Denial of Service (DDoS)}
DDoS attacks emerged as a more resilient and potent evolution of DoS, leveraging multiple sources, often part of a botnet, to overwhelm the target \cite{hoque2015botnet}. By decentralising the attack origin, DDoS campaigns are difficult to mitigate using source-based filtering techniques. Historic incidents such as the 2000 Mafiaboy attacks \cite{brooks2022ddos} and the 2016 Mirai botnet attack on Dyn DNS underscore the disruptive capacity of DDoS on critical infrastructure~\cite{antonakakis2017mirai}.

\subsection{Low-Rate Denial of Service (LDoS)}
LDoS attacks, often referred to as “shrew” attacks \cite{kuzmanovic2003shrew} or “slow” attacks, intentionally transmit minimal traffic volumes, typically below 1,000 packets per second \cite{xiang2011traceback} or within 10–20\% of baseline traffic levels \cite{zhijun2020lowrate}, in order to exploit vulnerabilities in timing mechanisms and queue management within network protocols \cite{kuzmanovic2006lowrate}. They aim to degrade performance without detection by traditional volumetric defences. Attacks like Slowloris \cite{damon2012slowloris} demonstrate how even low-bandwidth adversaries can inflict prolonged service degradation, especially in HTTP-layer contexts.

\subsection{Low-Rate Distributed Denial of Service (LDDoS)}
LDDoS attacks combine the stealth of LDoS with the distribution characteristics of DDoS \cite{siracusano2018lddos}. Multiple nodes cooperate to emit sporadic, low-volume traffic in synchronised patterns to evade anomaly detection systems. These attacks often mimic legitimate background traffic and can be highly effective in congesting cloud application layers while remaining statistically inconspicuous.

\subsection{Economic Denial of Sustainability (EDoS)}
EDoS attacks exploit the auto-scaling and pay-per-use models of cloud infrastructure by generating traffic patterns that trigger elastic resource provisioning \cite{hoff2008edos}, \cite{mukherjee2024resource}. Over time, this leads to excessive financial burden on the victim without necessarily breaching service availability. EDoS introduces a sustainability-focused denial vector, targeting the economic viability of the target system rather than its technical uptime \cite{sqalli2011edosshield}.

\subsection{Denial of Wallet (DoW)}
DoW attacks \cite{kelly2021dow} are an extension of EDoS, uniquely characterised by targeting serverless architecture or cloud functions such as AWS Lambda or Google Cloud Functions \cite{dorsett2025dowreview}. These attacks invoke short-lived, metered functions at scale, causing billing exhaustion. Unlike EDoS, DoW attacks operate with minimal resource impact but exploit billing granularity, making them particularly dangerous in cloud-native microservice architectures \cite{kelly2024downet}. While DoW attacks are not yet widely recognised within the industry \cite{dorsett2025ddos}, their prevalence is increasing, and they pose a significant risk of financial disruption.

\subsection{Distributed Denial of Wallet (DDoW)}
DDoW attacks distribute the invocation of serverless resources across multiple sources to increase scale and survivability. These attacks complicate attribution and defence by masking malicious invocation patterns as distributed application logic through attacks such as Blast, Continual Inconspicuous, and Background Chained attacks \cite{mileski2022ddow}. They represent the pinnacle of sustainability-denial evolution by uniting distribution, granularity, and economic intent into a single threat vector.

\section{Conditional Taxonomy Overview}
As shown in Table.~\ref{tab:condition_descriptions}, denial attacks can be classified based on six core conditions ranging from source distribution to infrastructure targeting. The baseline condition, C0, is the foundational criterion asserting that malicious requests, in any capacity, are sent to a target. This condition is universally true across all denial attacks and serves as the root requirement for any attack to be classified within this taxonomy. Condition C1 defines attacks in which the malicious requests originate from a single source. This reflects the traditional, non-distributed model of Denial-of-Service (DoS), typically characterised by brute-force resource exhaustion from an isolated attacker. In contrast, Condition C2 accounts for scenarios where malicious requests stem from multiple sources, most notably via botnets. This condition marks the transition to distributed architectures and is the hallmark of Distributed Denial-of-Service (DDoS) and its derivatives. It represents a key escalation in attack volume and detectability, as attacks in this class leverage large-scale botnets or coordinated nodes to overwhelm the target.

\begin{table}[h]
\centering
\caption{Denial Attack Conditions}
\renewcommand{\arraystretch}{1.2}
\begin{tabular}{|p{1.2cm}|p{6.3cm}|}
\hline
\textbf{Condition} & \textbf{Description} \\
\hline
C0 & Malicious requests, in any capacity, are sent to a target. \\
\hline
C1 & Denial of service in any capacity. \\
\hline
C2 & Requests originate from multiple sources. \\
\hline
C3 & Less than 1,000 packets or 10--20\% of normal traffic. \\
\hline
C4 & Targets scalable cloud infrastructure resources. \\
\hline
C5 & Targets serverless infrastructure. \\
\hline
\end{tabular}
\label{tab:condition_descriptions}
\end{table}

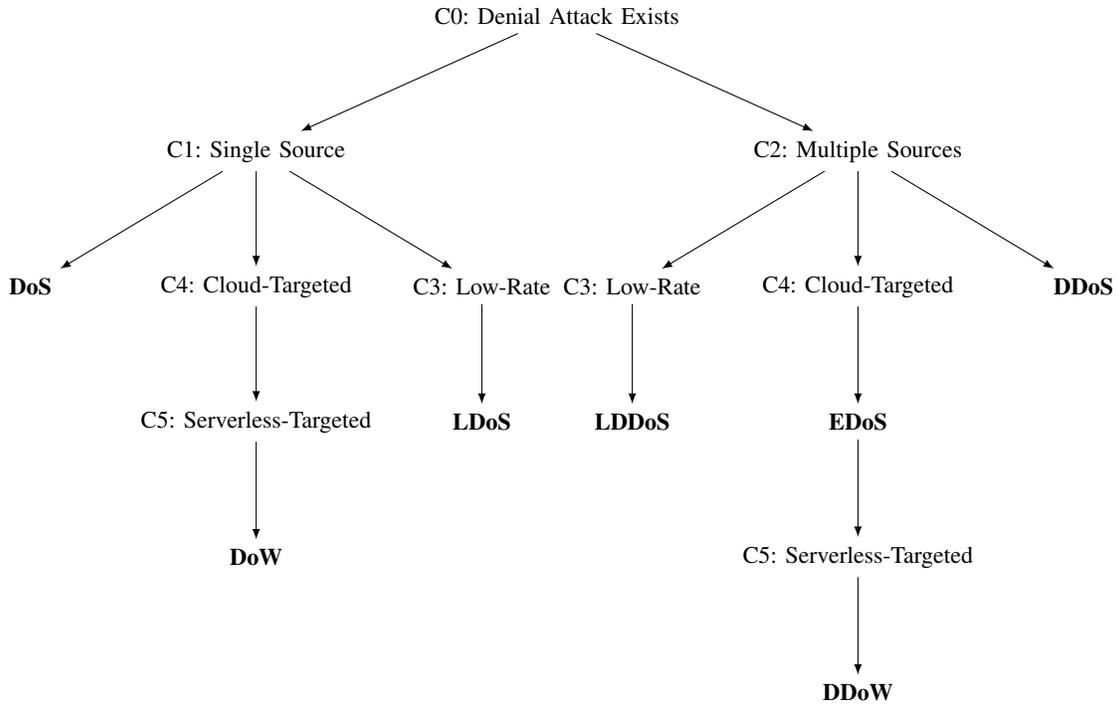
\begin{figure*}[t]
\centering
\scalebox{1}{
\begin{tikzpicture}[
  sibling distance=30pt,
  every node/.style = {font=\small},
  level distance=1.8cm,
  level 1/.style={sibling distance=80mm},
  level 2/.style={sibling distance=30mm},
  level 3/.style={sibling distance=20mm},
  edge from parent/.style={draw,-latex}
]

\node {C0: Denial Attack Exists}
  child { node {C1: Single Source}
    child { node {\textbf{DoS}} }
    child { node {C4: Cloud-Targeted}
      child { node {C5: Serverless-Targeted}
        child { node {\textbf{DoW}} }
      }
    }
    child { node {C3: Low-Rate}
      child { node {\textbf{LDoS}} }
    }
  }
  child { node {C2: Multiple Sources}
    child { node {C3: Low-Rate}
      child { node {\textbf{LDDoS}} }
    }
    child { node {C4: Cloud-Targeted}
      child { node {\textbf{EDoS}}
        child { node {C5: Serverless-Targeted}
          child { node {\textbf{DDoW}} }
        }
      }
    }
    child { node {\textbf{DDoS}} }
  };

\end{tikzpicture}
}
\caption{Denial Attack Tree Taxonomy Based on Conditions C0 to C5.}
\label{fig:conditional-tree-taxonomy}
\end{figure*}

Condition C3 introduces a subtler behavioural vector, characterising attacks that utilise low-and-slow strategies. Specifically, it defines attacks whose request volume is less than 1{,}000 packets or constitutes only 10--20\% of background traffic. This condition is particularly important in identifying Low-rate DoS (LDoS) and Low-rate Distributed DoS (LDDoS) attacks, which prioritise stealth and evasion over brute force. These attacks blend into normal traffic patterns, often bypassing threshold-based detection systems, and pose significant challenges to conventional monitoring frameworks. Condition C4 captures a shift in the target environment, indicating that the attack is directed at scalable cloud infrastructure. This is a critical advancement in the taxonomy, reflecting the emergence of Economic Denial-of-Sustainability (EDoS) and Denial of Wallet (DoW) attacks, both of which exploit the elasticity, pay-as-you-go, and consumption-based billing features of cloud computing platforms. Lastly, Condition C5 isolates attacks that specifically target serverless infrastructure, such as Function-as-a-Service (FaaS) environments. These environments operate on highly granular billing models, and as such, are uniquely susceptible to per-request financial exhaustion attacks. This condition is particularly prominent in the context of modern DoW and Distributed DoW (DDoW) attacks.\\

\begin{table*}[h]
    \centering
    \caption{Denial Attack Conditional Taxonomy}
    \renewcommand{\arraystretch}{1.35}
    \begin{tabular}{|p{7cm}|p{.8cm}|p{.8cm}|p{.8cm}|p{.8cm}|p{.8cm}|p{.8cm}|p{.8cm}|}
        \hline
        \textbf{Condition} & \textbf{DoS} & \textbf{DDoS} & \textbf{LDoS} & \textbf{LDDoS} & \textbf{EDoS} & \textbf{DoW} & \textbf{DDoW} \\
        \hline
        C0: Malicious requests, in any capacity, are sent to a target. & \checkmark & \checkmark & \checkmark & \checkmark & \checkmark & \checkmark & \checkmark \\
        \hline
        C1: Malicious requests are sent from only a single source. & \checkmark & & \checkmark & & & \checkmark & \\
        \hline
        C2: Malicious requests come from more than one source. & & \checkmark & & \checkmark & & & \checkmark \\
        \hline
        C3: Malicious requests make up less than 1,000 packets or 10--20\% of target background traffic. & & & \checkmark & \checkmark & & & \\
        \hline
        C4: Malicious requests target scalable cloud infrastructure resources. & & & & & \checkmark & \checkmark & \checkmark \\
        \hline
        C5: Malicious requests target serverless infrastructure. & & & & & & \checkmark & \checkmark \\
        \hline
    \end{tabular}
    \label{tab:denial-attack-conditional-taxonomy}
\end{table*}

Applying the taxonomy seen in Table ~\ref{tab:denial-attack-conditional-taxonomy} and Fig ~\ref{fig:conditional-tree-taxonomy} across well-known denial attack variants reveals the diverse operational mechanics that differentiate these classes. Traditional DoS attacks meet conditions C0 and C1, reflecting their single-source, brute-force origins. DDoS attacks satisfy C0 and C2, highlighting their reliance on distribution for impact amplification. LDoS and LDDoS share the C3 stealth condition but differ in source distribution—LDoS meeting C1, and LDDoS meeting C2. EDoS is distinguished by its explicit targeting of scalable cloud-based infrastructures (C4), setting it apart from volumetric unavailability goals. DoW represents a convergence of several conditions: C0, C1, C4, and C5. This configuration demonstrates a single-source attack targeting scalable cloud-native and serverless resources with the explicit intent of incurring operational costs without necessarily degrading service performance. Finally, DDoW emerges as the most conditionally rich attack class, encompassing C0, C2, C4, and C5. It represents a distributed and financially driven attack optimised for maximum undetectability and economic impact across serverless cloud environments.

\vspace{1em}
\section{Denial Attack Lattice}

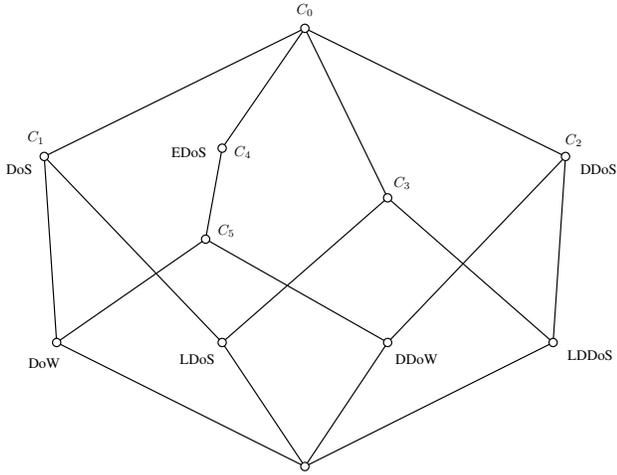
\begin{figure}[H]
\centering
\scalebox{0.55}{
\begin{tikzpicture}

\coordinate (bottom) at (0,0);
\coordinate (DoW) at (-6,3);
\coordinate (LDoS) at (-2,3);
\coordinate (DDoW) at (2,3);
\coordinate (LDDoS) at (6,3);
\coordinate (DoS) at (-6.3,7.5);
\coordinate (EDoS) at (-2,7.7);
\coordinate (C5) at (-2.4,5.5);
\coordinate (C3) at (2,6.5);
\coordinate (DDoS) at (6.3,7.5);
\coordinate (C0) at (0,10.6);

\draw [thick] (DoS) -- (C0) -- (DDoS) -- (LDDoS) -- (bottom) -- (DoW);
\draw [thick] (bottom) -- (LDoS) -- (C3) -- (LDDoS);
\draw [thick] (bottom) -- (DDoW) -- (DDoS);
\draw [thick] (DDoW) -- (C5) -- (EDoS) -- (C0);
\draw [thick] (C3) -- (C0);
\draw [thick] (LDoS) -- (DoS) -- (DoW) -- (C5);

\def\r{0.1}
\foreach \pt in {bottom,DoW,LDoS,DDoW,LDDoS,DoS,EDoS,C5,C3,DDoS,C0}
  \draw [thick,fill=white] (\pt) circle (\r);

\node at ($(DoW)+(-0.3,-0.5)$) {DoW};
\node at ($(DoS)+(-0.6,-0.3)$) {DoS};
\node at ($(DoS)+(-0.2,0.45)$) {$C_1$};
\node at ($(EDoS)+(-0.8,-0.1)$) {EDoS};
\node at ($(EDoS)+(0.5,-0.1)$) {$C_4$};
\node at ($(C0)+(0,0.45)$) {$C_0$};
\node at ($(DDoS)+(0.8,-0.3)$) {DDoS};
\node at ($(DDoS)+(0.2,0.4)$) {$C_2$};
\node at ($(C3)+(0.35,0.35)$) {$C_3$};
\node at ($(C5)+(0.5,0.2)$) {$C_5$};
\node at ($(DDoW)+(0.7,-0.4)$) {DDoW};
\node at ($(LDDoS)+(0.9,-0.3)$) {LDDoS};
\node at ($(LDoS)+(-0.6,-0.4)$) {LDoS};

\end{tikzpicture}
}
\caption{Lattice Diagram Representing the Conditional Hierarchy of Denial Attacks.}
\label{fig:denial-attack-lattice}
\end{figure}

The taxonomy’s lattice structure draws inspiration from order theory to model relationships between attacks based on cumulative condition satisfaction \cite{davey2002lattices}, \cite{koussas2021classification}. To capture escalation paths and partial orderings of attack complexity, Fig.~\ref{fig:denial-attack-lattice} presents a lattice diagram of conditionally dependent attack types.
The lattice diagram presented in this framework represents a formally structured, partially ordered set that models the hierarchical interrelationships between denial-of-service attack types and their associated conditional properties (C0-C5). Unlike the tabular taxonomy or Venn diagram, which offer horizontal or intersectional perspectives on classification, the lattice structure captures the vertical progression of conditional complexity and capability across denial attacks.

At the apex of the lattice is C0, the foundational condition shared by all denial attacks—namely, that malicious requests, in any form, are issued to a target. This root node is universally inherited by all descendant nodes in the lattice. Descending from C0, we observe bifurcations leading into C1 and C2, which represent the dichotomy between single-source and multi-source attack architectures. These conditions respectively define the transition into DoS and DDoS. Further refinement is enabled by the addition of C3, which encapsulates low-rate or stealth-based tactics. The path C0 $\rightarrow$ C1 $\rightarrow$ C3 yields LDoS, while C0 $\rightarrow$ C2 $\rightarrow$ C3 gives rise to LDDoS.\\

The lattice extends toward economic and infrastructural denial attacks. The path through C0 $\rightarrow$ C4 leads to EDoS. The addition of C5 (serverless targeting) yields the DoW and DDoW forms of attack. DoW arises from the sequential satisfaction of C0, C1, C4, and C5; DDoW emerges from C0, C2, C4, and C5. Each path within the lattice can be interpreted as a logical construction chain, where the satisfaction of a subset of conditions enables the classification of a denial attack type.

\section{Denial Attack Venn Diagram}
\begin{figure}[ht]
\centering
\includegraphics[width=0.368\textwidth]{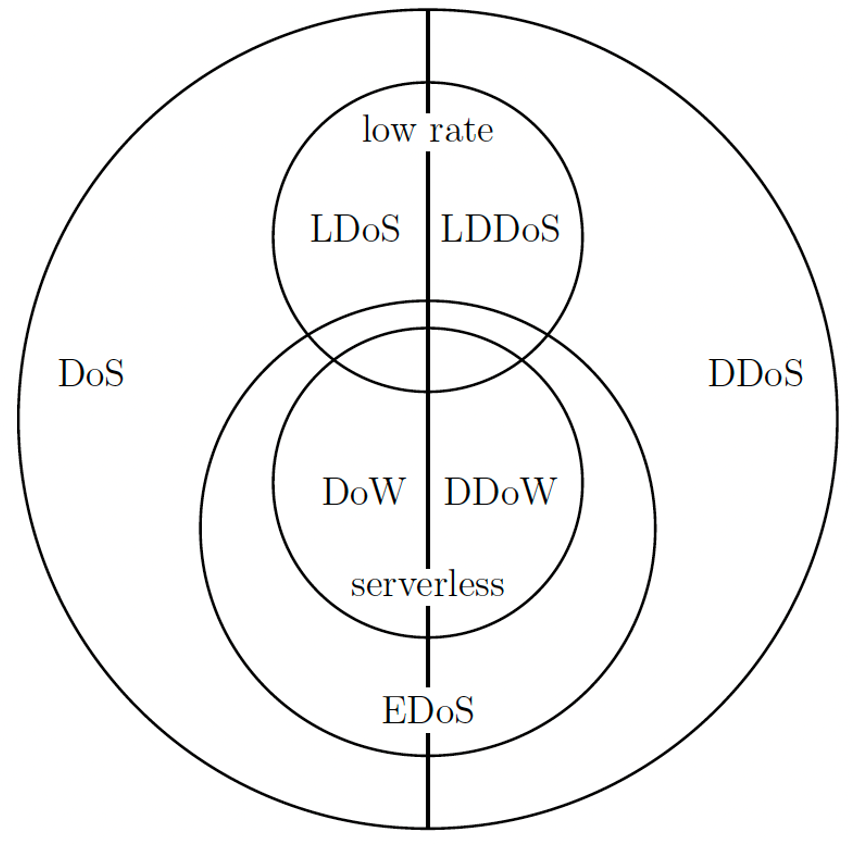}
\caption{Venn Diagram Representing the Overlap Between Availability and Sustainability-Based Denial Attacks.}
\label{fig:venn}
\end{figure}

The Venn diagram in Fig.~\ref{fig:venn} illustrates the intersection of Availability and Sustainability based denial attacks. It visualises the interrelationships among traditional and modern denial-of-service attack forms across availability and sustainability dimensions.

The vertical axis bisects the diagram into Availability-based and Sustainability-based Denial Attacks. DoS and DDoS reside in separate hemispheres. The "low-rate" domain, with LDoS and LDDoS, reflects stealth vector methodology. The "serverless" domain captures cost-based, cloud-native attacks, DoW and DDoW, linked to Function-as-a-Service (FaaS) billing models. EDoS overlaps with both hemispheres but roots more firmly in sustainability due to the financial impact attacks inflict.\\

This diagram reveals the continuum from brute-force availability attacks to stealthy, financially exploitative ones. It integrates both the mechanistic and strategic dimensions of denial attacks.

\section{Case Studies}

To demonstrate the practical applicability and analytical strength of the proposed conditional taxonomy, this section presents four case studies spanning both legacy and modern denial attack scenarios. These include a classical TCP SYN Flood (representing early volumetric attacks) which was pioneering, a simulated serverless billing attack (targeting financial sustainability), the 2016 Mirai botnet incident (a world-famous hallmark DDoS event), and the Slowloris attack (noted for its sophisticated and stealth footprint). Each case is evaluated against the six defined conditions (C0–C5) to illustrate the taxonomy’s expressiveness, backward compatibility with legacy denial attacks, and predictive alignment with evolving cloud and serverless attack vectors. Collectively, these examples highlight the taxonomy's robustness in characterising both availability- and sustainability-based attacks under a unified conditional schema.

\subsection{Real-World Case Study: TCP SYN Flood Attack}

The TCP SYN Flood in this case is a classical denial-of-service (DoS) attack from 1996 that pre-dated the use of Botnets, that exploited the TCP three-way handshake \cite{schuba1997tcp}. The attacker sends a rapid sequence of SYN packets to a server, initiating connections but never completing them with the final ACK. This results in half-open connections that consume memory and processing resources on the server. Over time, legitimate requests are dropped as the server’s backlog queue fills, leading to service degradation or full unavailability. This type of attack was demonstrated in the included analysis using a single host to flood a victim server, validating its impact on TCP throughput and latency.\\

Applying the proposed taxonomy, the TCP SYN Flood attack can be formally characterised by evaluating the attack conditions it satisfied.

\vspace{0.2cm}
\noindent\textbf{Conditions Met:} C0, C1

\vspace{0.2cm}
\noindent\textbf{Classification:} \textit{Denial of Service (DoS)}

\vspace{0.2cm}
\noindent\textbf{Justification:} The SYN Flood attack is a classical single-source DoS technique that satisfies the base condition (C0) and originates from a single host (C1). It does not use multiple sources (C2), does not employ low-rate stealth tactics (C3), and does not involve cloud (C4) or serverless (C5) infrastructures. This confirms its placement as a legacy availability-based denial attack. While it remains technically simple, its inclusion in the taxonomy showcases the framework's ability to encompass well-established threats alongside emerging ones, thus offering historical context within a unified classification structure.

\subsection{Simulated Case Study: Stealthy Serverless Billing Attack}

In this theoretical scenario, a serverless Function-as-a-Service (FaaS) backend (e.g., AWS Lambda, Google Cloud Functions, or Azure Functions) is exploited by an adversary orchestrating a distributed network of compromised edge devices. These devices each send low-volume, periodic API requests that mimic benign user behaviour — such as telemetry pings, health checks, or status updates — at carefully controlled intervals.\\

While each individual invocation is harmless and falls within the bounds of typical service usage, the aggregate effect across thousands of devices causes significant and sustained compute costs over time. This stealthy behaviour allows the attacker to evade detection by traditional volumetric DDoS mitigation systems, while achieving the desired goal: to exhaust the financial sustainability of the targeted cloud deployment.

\vspace{0.2cm}
\noindent\textbf{Conditions Met:} C0, C2, C3, C4, C5

\vspace{0.2cm}
\noindent\textbf{Classification:} \textit{Distributed Denial of Wallet (DDoW)}

\vspace{0.2cm}
\noindent\textbf{Justification:} The attack meets all five conditions associated with DDoW attacks. Its distributed nature (C2) and low-rate profile (C3) make it difficult to identify using anomaly detection. The exclusive targeting of serverless compute functions (C5) amplifies cost impact due to the pay-per-invocation billing model. Furthermore, as the infrastructure is cloud-based (C4), the attack leverages elastic scalability to scale the economic damage. This scenario exemplifies a sustainability-based denial attack that bypasses traditional uptime-oriented metrics and instead undermines operational viability.

\subsection{Real-World Case Study: The Mirai Botnet}

The 2016 Mirai botnet incident represents a landmark DDoS attack in Internet history. It leveraged vulnerable IoT devices—such as routers, IP cameras, and digital video recorders—using factory-default credentials \cite{antonakakis2017mirai}. Once compromised, these devices formed a massive botnet capable of launching unprecedented volumes of traffic. The most notable target was Dyn DNS, a critical DNS provider, whose outage cascaded across major web services including Twitter, Reddit, and Netflix.\\

Applying the proposed taxonomy, the Mirai attack can be formally characterised by evaluating the attack conditions it satisfied.

\vspace{0.2cm}
\noindent\textbf{Conditions Met:} C0, C2

\vspace{0.2cm}
\noindent\textbf{Classification:} \textit{Distributed Denial of Service (DDoS)}

\vspace{0.2cm}
\noindent\textbf{Justification:} Mirai was a volumetric, distributed denial-of-service attack that overwhelmed a critical internet-facing service. It was not stealthy (C3), did not abuse cloud-native elastic infrastructure (C4), and did not target serverless or billing-based exhaustion (C5). Though it targeted Dyn, a large-scale service provider operating on cloud infrastructure, it did not explicitly target scalability or elasticity; therefore, the Mirai attack does not satisfy Conditions C4 or C5. This demonstrates the taxonomy’s precision in distinguishing between traditional and cloud-era attacks. The exclusion of C4 and C5 reinforces that Mirai is firmly situated within the availability denial vector, without implications for financial sustainability. Its presence in the taxonomy nonetheless underscores the framework's ability to classify both legacy and modern threats under a unified structure.

\subsection{Legacy Case Study: Slowloris HTTP Denial Attack}

The Slowloris attack, introduced in 2009, remains one of the most notable examples of a low-rate denial-of-service technique that exploits the concurrency limits of web servers, particularly those running Apache. Instead of overwhelming the target with high traffic volumes, Slowloris initiates thousands of partial HTTP connections and intentionally delays completion. These half-open requests consume server resources by keeping threads or sockets active indefinitely.\\

Unlike volumetric attacks, Slowloris uses minimal bandwidth and does not require a distributed network to be effective, making it stealthy and resource-efficient. This attack has been observed in real-world targeting of media outlets, government websites, and activist platforms, often bypassing early forms of rate-based intrusion detection systems.

\vspace{0.2cm}
\noindent\textbf{Conditions Met:} C0 (malicious traffic), C1 (single origin), C3 (low-rate)

\vspace{0.2cm}
\noindent\textbf{Classification:} \textit{Low-rate Denial of Service (LDoS)}

\vspace{0.2cm}
\noindent\textbf{Justification:} The Slowloris attack satisfies three core conditions. It originates from a single device (C1) and sends incomplete packets in a prolonged, time-delayed fashion (C3), making it a clear match for low-rate classification. Despite its age, the attack remains relevant in constrained environments or where bandwidth is limited. It does not leverage cloud-based or serverless infrastructure (C4, C5), aligning it strictly within the availability-based denial spectrum and demonstrating the backward compatibility of the proposed taxonomy.

\subsection{Summary}

These case studies validate the adaptability of the proposed taxonomy. From legacy attacks like Mirai (DDoS) and Slowloris (LDoS) to modern cloud-native threats such as serverless billing abuse, the taxonomy consistently distinguishes between availability and sustainability attack vectors, enabling specific classification of denial attacks, as seen in Table \ref{tab:case-summary}.

\begin{table}[htbp]
\centering
\caption{Taxonomy-Based Classification of Case Studies}
\label{tab:case-summary}
\resizebox{\columnwidth}{!}{%
\begin{tabular}{|l|c|c|c|c|c|c|l|}
\hline
\textbf{Attack Type} & \textbf{C0} & \textbf{C1} & \textbf{C2} & \textbf{C3} & \textbf{C4} & \textbf{C5} & \textbf{Classification} \\
\hline
TCP SYN Flood & \cmark & \cmark & \xmark & \xmark & \xmark & \xmark & DoS \\
\hline
Serverless Billing Attack & \cmark & \xmark & \cmark & \cmark & \cmark & \cmark & DDoW \\
\hline
Mirai Botnet & \cmark & \xmark & \cmark & \xmark & \xmark & \xmark & DDoS \\
\hline
Slowloris & \cmark & \cmark & \xmark & \cmark & \xmark & \xmark & LDoS \\
\hline
\end{tabular}
}
\end{table}

\section{Limitations and Future Work}

While the proposed conditional taxonomy offers a structured and formal framework to classify denial attacks, several limitations must be acknowledged.\\

First, the model assumes discrete binary conditions (met or unmet), which may oversimplify the nuanced nature of real-world attacks. Some conditions, such as C3 (stealth) or C4 (cloud targeting), may present on a spectrum rather than a binary scale, depending on the attack strategy or deployment context, with Mirari Botnet as a perfect example.\\

Second, this taxonomy is currently static and expert-defined. It does not incorporate adaptive mechanisms or machine learning classifiers that could dynamically evolve the taxonomy as new denial attack types emerge. Future research may benefit from hybrid models that integrate static taxonomies with anomaly-based or learning-based classifiers to improve real-time applicability.\\

Third, while the framework is extensible to sustainability-based attacks such as DoW, empirical validation across a broader set of attacks, including those not yet categorised in academic literature, is still limited. The case studies included herein offer initial validation but do not cover the full spectrum of contemporary cloud-native threats.\\

Finally, serverless infrastructures continue to evolve rapidly, introducing new event-driven architectures, edge computing layers, and billing models. These emerging attributes may alter the relevance or thresholds of conditions C4 and C5, requiring future refinement or extension of the taxonomy.\\

\textbf{Future work} will aim to:
\begin{itemize}
    \item Incorporate spectrum-based (fuzzy) conditions, rather than assume binary.
    \item Integrate the taxonomy with human-socio conditions.
    \item Apply the taxonomy to industry datasets.
    \item Expand the taxonomy with additional conditions (e.g., intent, persistence, recovery time).
\end{itemize}

\subsection{Proposed Validation Strategy}

To address empirical gaps, future work will involve the taxonomy being tested against attack datasets and synthetic denial attack simulations of both availability and sustainability based vectors to assess classification accuracy. 

\subsection{Tooling and Practical Integration}

Integrating the taxonomy into security operations workflows may significantly enhance denial attack classification. Future iterations may explore how conditions C0–C5 can be mapped to billing alerts and cloud-native observability tools. This can support both proactive detection and reactive post-incident forensic classification of denial tactics.

\section{Conclusion}
The proposed taxonomy, lattice, and Venn diagram offer a comprehensive, logically grounded classification system for denial-of-service attacks. It unites legacy and modern variants under a condition-driven schema, highlights hybrid threats, and aids both theoretical exploration and real-world threat modeling in cloud and serverless environments. This condition-driven perspective not only clarifies existing threat models but also lays a scalable foundation for understanding future denial tactics in increasingly complex, financially-driven, and serverless cloud ecosystems.

\bibliographystyle{IEEEtran}

\end{document}